\input harvmac

\def\Title#1#2{\rightline{#1}\ifx\answ\bigans\nopagenumbers\pageno0
\else\pageno1\vskip.5in\fi \centerline{\titlefont #2}\vskip .3in}

\font\caps=cmcsc10

\noblackbox
\parskip=1.5mm

  
\def\npb#1#2#3{{\it Nucl. Phys.} {\bf B#1} (#2) #3 }
\def\plb#1#2#3{{\it Phys. Lett.} {\bf B#1} (#2) #3 }
\def\prd#1#2#3{{\it Phys. Rev. } {\bf D#1} (#2) #3 }

\def\ijmpa#1#2#3{{\it Int. J. Mod. Phys.} {\bf A#1} (#2) #3 }

\def\pr#1#2#3{{\it Phys. Rep. } {\bf #1} (#2) #3 }

\def\bb#1{{\tt hep-th/#1}}


\def\CL{{\cal L}}

\def\CN{{\cal N}}


\def\dj{\hbox{d\kern-0.347em \vrule width 0.3em height 1.252ex depth
-1.21ex \kern 0.051em}}

\def\ket{\rangle}
\def\bra{\langle}

\def\pt{\partial}

\def\epb{\overline \varepsilon}   
\def\ep{\varepsilon}    
\def\zb{\overline z} 
\lref\regk{S. Elitzur, A. Giveon and D. Kutasov, 
 \plb{400}{1997}{269,} \bb{9702014.}}
\lref\regkrs{S. Elitzur,
 A. Giveon, D. Kutasov, E. Rabinovici and A. Schwimmer, 
{\it ``Brane Dynamics and N=1 Supersymmetric Gauge Theory",}
\bb{9704104.}}   
\lref\rvaf{M. Bershadsky, A. Johansen, T. Pantev, V. Sadov and
C. Vafa, {\it ``F-theory, Geometric Engineering and $N=1$ dualities",}
\bb{9612052\semi} 
C. Vafa and B. Zwiebach, {\it ``$N=1$ Dualities of
$SO$  and $Usp$ Gauge Theories and T-Duality of String Theory",}
\bb{9701015\semi}
C. Vafa and H. Ooguri, {\it ``Geometry of $N=1$ dualities
in Four Dimensions",} 
\bb{9702180.}}
\lref\rhw{A. Hanany and E. Witten, \npb{492}{1997}{152,} \bb{9611230.}} 
\lref\rdeff{P.C. Argyres, M.R. Plesser and N. Seiberg, \npb{471}{1996}
{159,}  \bb{9603042.}} 
\lref\rpol{J. Polchinski, {\it ``TASI lectures on D-branes'',} 
\bb{9611050.}}
\lref\rdm{M.R. Douglas and G. Moore, {\it ``D-branes, Quivers and 
ALE Instantons'',} \bb{9603167}.}
\lref\rwinst{E. Witten, \npb{460}{1996}{335,} \bb{9510135\semi} 
\npb{460}{1996}{541,} \bb{9511030.}}
\lref\rdoug{M.R. Douglas, {\it ``Branes within Branes'',} \bb{9512077\semi} 
{\it ``Gauge Fields and D-branes'',} \bb{9604198.}}
\lref\rvafi{C. Vafa, \npb{463}{1996}{435,} \bb{9512078.}}
\lref\rahh{O. Aharony, A. Hanany, {\it ``Branes, Superpotentials
and Superconformal Fixed Points",} 
\bb{9704170.}}
\lref\rbdh{J.H. Brodie and A. Hanany, {\it `` Type-IIA Superstrings,
Chiral Symmetry and $N=1$ 4D Gauge Theory Duality",} 
\bb{9704043.}} 
\lref\rahz{A. Hanany and A. Zaffaroni, {\it ``Chiral Symmetry from
Type-IIA Branes",} 
\bb{9706047.}}
\lref\rclf{N. Evans, C.V. Johnson and A.D. Shapere, {\it ``Orientifolds,
Branes, and Duality of 4D Gauge Theories",} \bb{9703210.}} 
\lref\rbarb{J.L.F. Barb\'on, \plb{402}{1997}{59,} \bb{9703051.}} 
\lref\rberk{J. de Boer, K. Hori, H. Ooguri, Y. Oz and Z. Yin, \npb{493} 
{1997}{148,} \bb{9702154.}} 
\lref\rtheis{A. Brandhuber, J. Sonnenschein, S. Theisen and S. 
Yankielowicz, {\it ``Brane Configurations and 4D Field Theory Dualities",} 
\bb{9704044.}} 
\lref\rwittenM{E. Witten, {\it ``Solutions Of Four-Dimensional Field
Theories via M Theory",} \bb{9703166.}} 
\lref\rmberk{K. Hori, H. Ooguri and Y. Oz, {\it ``Strong Coupling
Dynamics of Four-Dimensional $N=1$ Gauge Theories from M Theory
Fivebrane",} \bb{9706082.}}
\lref\rkapl{A. Brandhuber, N. Itzhaki, V. Kaplunovsky, J. Sonnenschein
and S. Yankielowicz, {\it ``Comments on the M Theory Approach to
$N=1$ $SQCD$ and Brane Dynamics",} \bb{9706127.}} 
\lref\rrut{O. Aharony, A. Hanany, K. Intriligator, N. Seiberg and
M.J. Strassler, {\it ``Aspects of $\CN =2$ Supersymmetric Gauge
Theories in Three Dimensions",} \bb{9703110.}} 
\lref\rwitqcd{E. Witten, {\it ``Branes And The Dynamics Of QCD",} 
\bb{9706109.}} 
\lref\rmqcd{A. Hanany, M.J. Strassler and A. Zaffaroni, {\it ``Confinement
and Strings in MQCD",} \bb{9707244.}} 
\lref\raffl{I. Affleck, \npb{191}{1981}{429.}} 
\lref\rtof{G. 't Hooft, \prd{14}{1976}{3432.}}
\lref\ritep{V.A. Novikov, M.A. Shifman, A.I. Vainshtein, V.B. Voloshin
and V.I. Zakharov, \npb{229}{1983}{394\semi} 
V.A. Novikov, M.A. Shifman, A.I. Vainshtein and V.I. Zakharov, \npb{260}
{1985}{157.}}
\lref\rrev{D. Amati, K. Konishi, Y. Meurice, G. Rossi and G. Veneziano,
\pr {162}{1988}{169.}}                            
\lref\rdorey{N. Dorey, V.V. Khoze and M.P. Mattis, \prd {54}{1996}{2921,} 
\bb{9603136\semi} 
\prd{54}{1996}{7832,}
\bb{9607202.}}
\lref\rgomez{C. G\'omez and R. Hern\'andez, {\it ``M and F theory instantons,
N=1 supersymmetry and fractional topological charges'',} \bb{9701150.}}
\lref\ryi{K. Lee and P. Yi, {\it ``Monopoles and instantons on partially
compactified D-branes'',} \bb{9702107.}}
\lref\rangles{M. Berkooz, M.R. Douglas and R.G. Leigh, \npb{480}{1996}{265,}
\bb{9606135.}}  
\lref\rads{I. Affleck, M. Dine and N. Seiberg, \npb{241}{1984}{493.}}
\lref\rlands{K. Landsteiner, E. Lopez and D.A. Lowe, {\it ``$N=2$ 
Supersymmetric Gauge Theories, Branes and Orientifolds",} \bb{9705199.}}
\lref\rbrand{A. Brandhuber, J. Sonnenschein, S. Theisen and 
S. Yankielowicz, {\it ``M Theory and Seiberg--Witten Curves: Orthogonal
and Symplectic Groups",} \bb{9705232.}} 
\lref\rwihiggs{E. Witten, {\it ``On The Conformal Field Theory Of The Higgs
Branch'',} \bb{9707093.}} 
\lref\revans{N. Evans and M. Schwetz,
{\it ``The Field Theory of Non-Supersymmetric Brane Configurations'',} 
\bb{9708122.}}
\lref\rozdb{J. de Boer and Y. Oz,
{\it `` Monopole Condensation and Confining Phase of N=1 Gauge
Theories via M-theory Five-Brane'',}
\bb{9708044.}}
\lref\rhenn{A. Fayyazuddin and M. Spalinski, {\it
``The Seiberg--Witten differential from M-theory,"} \bb{9706087\semi} 
M. Henningson and P. Yi, {\it ``Four-Dimensional
BPS Spectra via M-Theory,''}
\bb{9707251\semi} 
A. Mikhailov, {\it ``BPS States and Minimal Surfaces,''} 
\bb{9708068.}}
\lref\rvafler{A. Klemm, W. Lerche, P. Mayr, C. Vafa and N. Warner,
\npb{477}{1996}{746,} \bb{9604034\semi}
M. Bershadsky, A. Johansen, T. Pantev, V. Sadov and
C. Vafa, {\it ``F-theory, Geometric Engineering and $N=1$ dualities",}
\bb{9612052.}}
\lref\rus{J.L.F. Barb\'on and A. Pasquinucci, {\it ``D0-branes,
Constrained Instantons and D=4 Super Yang--Mills Theories'',}
\bb{9708041.}}
\lref\rtheta{N. Evans, S. Hsu and M. Schwetz, \npb{484}{1997}{124,}
\bb{9608135,~} \plb{404}{1997}{77,} \bb{9703197\semi}
K. Konishi and M. Di Pierro, \plb{388}{1996}{90,} \bb{9605178\semi}
K. Konishi, \plb{392}{1997}{101,} \bb{9609021.}}
\lref\ryale{N. Evans, S. Hsu and M. Schwetz, \plb{355}{1995}{475,}
\bb{9503186\semi} N. Evans, S. Hsu, M. Schwetz and S.B. Selipsky,
\npb{456}{1995}{205,} \bb{9508002\semi} 
O. Aharony, J. Sonnenschein, M.E. Peskin and S. Yankielowicz,
\prd{52}{1995}{6157,}  \bb{9507013.}}
\lref\rlag{L. Alvarez-Gaum\'e, J. Distler, C. Kounnas and 
M. Mari\~no, \ijmpa{11}{1996}{4745,} \bb{9604004\semi}
L. Alvarez-Gaum\'e and M. Mari\~no, \ijmpa{12}{1997}{975,} \bb{9606191\semi}
L. Alvarez-Gaum\'e, M. Mari\~no and F. Zamora, {\it ``Softly Broken 
N=2 QCD with Massive Quark Hypermultiplets, I'',} \bb{9703072;} 
{\it ``Softly Broken N=2 QCD with Massive Quark Hypermultiplets, II'',}
\bb{9707017.} }
\lref\rwtheta{E. Witten, {\it Ann. Phys.} (N.Y.) 128 (1980) 363.} 
\lref\rtofi{G. `t Hooft, {\it Phys. Scr.} {\bf 24} (1981) 841; 
{\bf 25} (1981) 133; \npb {190}{1981}{455.}} 
\lref\rsmilga{A. Smilga, \prd{49}{1994}{6836.}}  
\lref\rdashen{R.F. Dashen, \prd{3}{1971}{1879.}}



\line{\hfill CERN-TH/97-308}
\line{\hfill KUL-TF-97/28}
\line{\hfill {\tt hep-th/9711030}}
\vskip 1.2cm

\Title{\vbox{\baselineskip 12pt\hbox{}
 }}
{\vbox {\centerline{Softly Broken MQCD and the Theta Angle}
}}

\vskip0.6cm

\centerline{$\quad$ {\caps J. L. F. Barb\'on~$^1$ and A. Pasquinucci~$^2$
 }}
\vskip0.8cm

\centerline{{\sl $^1$ Theory Division, CERN}}
\centerline{{\sl 1211 Geneva 23, Switzerland}}
\centerline{{\tt barbon@mail.cern.ch}}

\vskip0.3cm

\centerline{{\sl  $^2$ Instituut voor theoretische fysica, K.U.\
Leuven}}
\centerline{{\sl  Celestijnenlaan 200D }}
\centerline{{\sl  B-3001 Leuven, Belgium  }}

\vskip 1.0in

\noindent{\bf Abstract:} We consider a family of non-supersymmetric 
 MQCD five-brane configurations  introduced by Witten, 
and  discuss the dependence of the curves
on the microscopic theta angle and its relation with CP.
We find evidence for a non-trivial spectral flow of the curves (vacua)
and for the level-crossing of adjacent curves at a particular value
of the theta angle, with  spontaneous  breaking of CP symmetry,  
 providing an MQCD analogue of the phase transitions in theta proposed
by 't Hooft.  

%
%
%


\Date{November 1997}



The modelling of gauge dynamics via brane configurations of weakly coupled
string theory \refs{\rhw,\regk,\regkrs,\rberk}, or M-theory 
\refs{\rwittenM,\rwitqcd,\rmqcd,\rmberk,\rkapl},
provides a geometrical interpretation of various field 
theoretical strong coupling
phenomena. In some cases, the geometrical viewpoint  can be used
to discover new exact solutions \refs\rwittenM\ of $N=2$ theories. In the
context of $N=1$ theories, it provides a semiclassical approach to
such thorny problems as confinement and chiral symmetry breaking 
\refs{\rwitqcd,\rmqcd}. 
These constructions are similar in spirit to previous work on
geometric engineering of gauge theories (see for example \refs\rvafler).
Here the role of a
non-trivial string compactification is played by some complicated
brane configuration sitting in flat space at
small string coupling, and carefully adjusted to provide a weakly coupled
supersymmetric Yang--Mills theory at intermediate scales. Since the
gauge theory lives on the branes world-volume, one can relate many
classical and semiclassical features of gauge theories to some properties
of brane dynamics. The real power of the method arises when taking the
strong coupling limit of the brane configuration. If a strong coupling dual
is available, one can describe many non-perturbative, infrared properties
of the gauge theory, just by reading the tree-level data of the dual brane 
configuration.  

In four-dimensional models one uses the duality between type IIA strings
and M-theory \refs\rwittenM.
 Here one maps the type IIA brane configuration to
a single smooth five-brane, whose world-volume is appropriately embedded
in the flat eleven-dimensional background, as a product $M_4 \times \Sigma$,
with $M_4$ the four-dimensional Minkowski space, and $\Sigma$ a holomorphic
curve with respect to a given complex structure of the background.
  In general, the detailed physics of the resulting M-theory 
model differs  from the Yang--Mills theory we are interested in; however,  
to the extent that some observables are protected by supersymmetry, we
can calculate them in the deformed strong coupling model. This is the 
case, for example, of all holomorphic 
 quantities determined by the Seiberg--Witten
curve in $N=2$ models \refs\rwittenM, including BPS spectra \refs\rhenn.
 For $N=1$ models
 the agreement is only qualitative in principle,
but some observables can be accurately matched, like superpotentials and
gaugino condensates \refs{\rwitqcd,\rmberk}.

 In the absence of some unbroken supersymmetry on the
world-volume, we cannot use holomorphy to accurately match observables, and
a working assumption must be made that no phase transitions occur in the
way to strong coupling. Still, qualitative features based on topological
properties can be established  \refs{\rwitqcd,\rmqcd}.

In this note we  study some qualitative features of a class
of non-supersymmetric brane configurations. In particular, using 
selection rules provided by the symmetries,  we present
evidence for a non-trivial structure of the spectral flow with respect
to the microscopic theta parameter,  $\theta$, with the associated
phase transitions as suggested by 't Hooft in the field theory
context \refs\rtofi.
Reviews of the relevant facts regarding the so-called ``theta
puzzle", and its connection with the Veneziano-Witten formula, can be
found in \refs\rwtheta, \refs\rsmilga. 
 For related discussions of non-supersymmetric brane configurations,
see \refs{\rkapl,\rtheis,\revans}.    

In the non-supersymmetric case, 
the criterion of perturbative stability of the five-brane
background becomes simply
the extremality with respect to local deformations of the embedding.
Namely, for a given set of asymptotic boundary conditions, the five-brane
world-volume 
is embedded in eleven-dimensional space as $M_4 \times \Sigma$, 
where now $\Sigma$ is a two-dimensional
surface of ``minimal area''. This surface being infinite in the flat
eleven-dimensional metric, the requirement is really stability with
respect to small local deformations.

Part of the asymptotic data are easily identifiable in the 
weak coupling limit of the brane configuration, in terms of intersecting
branes of type IIA string theory. Starting with a type IIA 
$N=2$ configuration
consisting of a pair of parallel NS5-branes on the $x^4, x^5$ plane,
and a set of  $n$ D4-branes stretched in
between along the $x^6$ direction, we shall
consider arbitrary rotations of one of the NS5-branes
into the four-dimensional space parametrized by $x^4, x^5, x^8, x^9$.
The corresponding rotations form a coset ${SO(4) \over SO(2) \times
SO(2)}$, parametrized by four angles. In the complex structure
$z=x^4 +ix^8$, $z'=x^5 +ix^9$, a particular $SU(2)$ rotation of the
form $e^{i\alpha} z$, $e^{-i\alpha} z'$ preserves $N=1$ supersymmetry
in the $M_4$ space-time \refs\rbarb, the angle of rotation being related to 
the mass  that splits  
 the $N=2$ vector multiplet into a massless $N=1$
vector multiplet and a massive chiral multiplet in the adjoint representation.
The rest of the parameters should be associated to supersymmetry-breaking
masses for the gauginos, and supersymmetry-breaking mass splittings
between the adjoint scalars and matter fermions. On the other
hand, the bare Yang--Mills coupling is related to the separation
of the NS5-branes along $x^6$. One coupling parameter, which is not
easily identifiable in the type IIA picture, is the bare theta angle.
One needs to switch on non-perturbative corrections to the classical
brane geometry in order to be sensitive to the value of theta \refs\rus.
For applications involving $N=2$ or $N=1$ supersymmetric backgrounds,
the actual value of the microscopic theta parameter is irrelevant,
since there are anomalous $U(1)$ symmetries in those cases.

Witten has shown how these branes configurations can be lifted to M-theory.
In the representation of the five-brane world-volume as $M_4 \times \Sigma$,
one first identifies $\Sigma$ with the complex $z$ plane, with $z=0$ and
$z=\infty$ associated to the two asymptotic regions of the NS5-branes. 
 Then, 
a ``minimal area'' embedding ${\vec X} (z,\zb)$ is characterized by
harmonic functions with 
a vanishing two-dimensional energy-momentum tensor~\foot{The mixed
components $T_{z\zb}$ vanish because of the  two-dimensional classical
Weyl invariance.}   
\eqn\emt{
T_{zz} = g_{ij} \pt_z X^i \pt_z X^j =0,}
where $g_{ij}$ is the background metric in eleven dimensions,  
 in M-theory units: 
\eqn\bm{
ds^2 = \sum_{i,j=0}^{9} \eta_{ij} dx^i dx^j  
  + R^2 (dx^{10})^2 . }  
Witten configurations in parametric form are, in the ${\vec X} = 
(x^4, x^5, x^8, x^9)$ space:  
\eqn\wc
{ {\vec X} (z,\zb) = {\rm Re}({\vec p}\, z + {\vec q}\, z^{-1}).}
The complex vectors ${\vec p}, {\vec q}$ define the asymptotic
orientation of the NS5-branes in the weak coupling limit: the region
with $z\rightarrow 0$ corresponds to the ``left" NS5-brane, while the
region with $z\rightarrow \infty$ leads to the ``right" NS5-brane. In
addition, the  
five-brane configuration wraps $n$ times 
the compact circle of radius $R$.
Choosing an angular variable in which the five-brane wraps rigidly
we have
\eqn\exten{x^{10} = -n\,{\rm Im} \,({\rm log}\,z).}
Finally, the profile of the five-brane in the $x^6$ direction is
parametrized as
\eqn\exsix{x^6 = -R\,n \,{\rm Re}\, (c\, {\rm log}\, z)  }  
with  $c$ a real constant. 

The leading terms as $z\rightarrow 0$ and $z\rightarrow \infty$ of
the vacuum equations \emt\ are  ${\vec p}^{\,2} =0$ and
${\vec q}^{\,2} =0$, respectively.
 In addition, a subleading term in \emt\ relates  $c$ 
 to the asymptotic vectors:    
\eqn\cfix{ {\vec p}\cdot {\vec q} = {R^2 n^2 \over 2} (c^2 -1). }
Notice that \cfix\ makes sense for complex $c$, since the vectors
${\vec p}$ and ${\vec q}$ are complex. However, if ${\rm Im}(c) \neq 0$
in \exsix, the embedding becomes multivalued in the $x^6$ direction.
Since $x^6$ is non-compact, a continuous embedding of $\Sigma$ requires
${\rm Im}(c) = 0$. For notational convenience, we find it useful to
work with the extended family of embeddings with a general complex $c$.
The reader should consider, however, that only the ${\rm Im}(c) =0$
family has a natural physical interpretation. 
    
Using the isometries of the metric \bm\ we may rotate the left
NS5-brane, given by ${\vec X} = {\rm Re} ({\vec p}\, z)$, 
 such that it lies along the $x^4, x^5$ plane. Furthermore,
rescaling $z$, we can bring ${\rm Re}\, {\vec p}$ and ${\rm Im}\,{\vec p}$
to unit vectors, so that ${\vec p}^{\,2} =0$ implies ${\rm Re}\,{\vec p}\,
\cdot\,{\rm Im}\,{\vec p} =0$. 
Up to parity transformations in the $x^4, x^5$ plane,
we may then completely fix the first vector  to, say,
 ${\vec p} = (1, -i, 0,0)$. 

A convenient parametrization of supersymmetry breaking arises when using
the complex structure $v= x^4 + ix^5$, $w= x^8 +ix^9$, $ t=e^{-s} $, 
$s=R^{-1} x^6  
+ix^{10}$, in terms of which the relevant part of the background metric
becomes
\eqn\bmm{ ds^2 = |dv|^2 + |dw|^2 + R^2 {|dt|^2 \over |t|^2} .} 
In these variables,  a subgroup $U(2)\times U(1) $ of the compact  
``internal" isometry group $O(5)\times U(1)$ is manifest in terms of
complex rotations  of $(v,w)$, and phase redefinitions  
of $t$. In addition we also have the discrete complex conjugation
symmetries of all variables, and inversions of $t$.      
If we parametrize the ``right" NS5-brane vector ${\vec q}$ as 
\eqn\qu{ {\vec q} = (\eta + \ep, -i\eta +i\ep , \zeta +\lambda, 
-i\zeta +i\lambda )  , } 
 the null condition ${\vec q}^{\,2} =0$ translates into   
\eqn\nul{\eta\ep + \zeta\lambda =0.}   
Then, the most general $N=0$ curve takes the simple form:
\eqn\curve{\eqalign{ v=& z + {\eta \over z} + {\epb \over \zb} \cr 
w =& {\zeta \over z} + {{\overline \lambda} \over \zb} \cr
t =& z^{n(c+1)/2} \, \zb^{n({\overline c} -1)/2} }   }   
and is specified by three complex parameters (for example $\eta$, $\zeta$
and $\ep$) since the remaining equation of motion, eq.\ \cfix, relates the 
constant $c$ with $\ep$:    
\eqn\epf{ {\vec p}\cdot {\vec q} = 2\,\ep = {R^2 n^2 \over 2} (c^2 -1)} 
In particular, reality of $c$ implies reality of $\ep$.\foot{The sign
ambiguity of $c$ as a solution of eq.\
\cfix\ is equivalent to the symmetry $t\rightarrow {\overline t}^{-1}$,
i.e.\ $x^6 \rightarrow -x^6$.}
The required reality condition on $\ep$ or $c$ raises an interesting
point. Notice that, from the point of view of the IIA brane configuration,
there are four angles parametrizing the most general rotation within
the $(v,w)$ hyperplane. This would correspond to the four degrees of
freedom contained in ${\vec q}$, after subtracting the two degrees
of freedom characterizing unrotated 
$N=2$ curves (the Yang--Mills coupling and
theta angle),  
 and the constraints from the leading field equation
at infinity: ${\vec q}^2 =0$.  However, the subleading equation 
\epf, together with the reality condition on $c$, imposes one extra
constraint on ${\vec q}$, and we find that only a subclass of the
rotated IIA brane configurations, depending on three real parameters,
 can be lifted into M-theory in a
smooth way.

Among the configurations described by the curve in  eq.\ \curve, 
the supersymmetric ones are holomorphic embeddings and,
accordingly, correspond to $\ep =0$,   
which in turn implies $c^2 =1$. 

The ``field equations" \nul\ and \epf\ have an obvious symmetry under
the interchange of $\lambda$ and $\zeta$. This symmetry is translated
into the curve \curve\ as the isometry $w\rightarrow {\overline w}$, and
implies that, for fixed $\eta$ and fixed supersymmetry-breaking parameter
$\ep$, we can restrict the values of $\zeta$ as $|\zeta| \ge \sqrt{|\eta
\ep|}$.  This is analogous to the effect of T-duality on the restriction
of the moduli space of a string compactification.  

Depending on the values of the (complex)
parameters $\eta$, $\ep$, $\zeta$ and $\lambda$ subject to eq.\ \nul, and
up to the replacements $\lambda \rightarrow \zeta$ and 
$w \rightarrow {\overline w}$, the curve in eq.\ \curve\ has 
the following limiting regimes:  
\item{{\sl i}.}{The $N=2$ curve at the singular
points where $n-1$ dyons become massless:
\eqn\cuno{ v = z + {\eta \over z}\ ,\qquad\qquad\quad
 w = 0\ ,\qquad\qquad\quad  t= z^n .}    
This is the case in which $\ep=\lambda=\zeta=0$ (i.e.\ $\vec{q}=\eta\vec{p}$).}
\item{{\sl ii}.}{The generic $N=1$ curve, that is MQCD  in presence 
of an adjoint chiral multiplet of mass proportional to
 $\mu=\zeta/\eta$:
\eqn\cdos{ v = z + {\eta \over z}\ ,\qquad\qquad\quad 
 w = {\zeta\over z} \ ,\qquad\qquad\quad
t= z^n } 
(the $N=2$ limit corresponds to $\mu\rightarrow 0$ at $\eta$ fixed).}
\item{{\sl iii}.}{The $N=1$ MQCD curve
\eqn\ctres{ v = z \ ,\qquad\qquad\quad
 w = {\zeta\over z} \ ,\qquad\qquad\quad
t= z^n ,}
which corresponds to the limit $\mu\rightarrow\infty$ at $\zeta$ fixed
of the previous configuration.}
%
%
\item{{\sl iv}.}{The $N=1$ MQCD curve softly broken to $N=0$ supersymmetry by 
the breaking term $\epb /\zb$
\eqn\ccin{ v = z + {\epb \over \zb}\ ,\qquad\qquad
 w = {\zeta\over z}\ ,\qquad\qquad
t= z^{n(c+1)/2} \zb^{n(\bar{c}-1)/2},  }
where $\lambda=\eta=0$.}

\vskip4pt
 
Notice that the two-dimensional surface $\Sigma$,  described by the
curve \curve,  is embedded in the $N=2$ case 
in a four-dimensional space
spanned by $(x_4,x_5,x_6,x_{10})$. In the $N=1$ case $\Sigma$ is embedded in
a six-dimensional space $(x_4,x_5,x_6,x_8,x_9,x_{10})$. Going to  $N=0$,  
since we lose holomorphicity, there appear more cases. For 
generic values of the parameters in the curve (with $\ep\neq 0$), $\Sigma$
is embedded in the same six-dimensional space as the $N=1$ case. But if we
choose $c=0$ (i.e.\  $\ep=R^2n^2/4$) then $\Sigma$ is embedded in a
five-dimensional space spanned by $(x_4,x_5,x_8,x_9,x_{10})$. Instead, for
$\eta=\lambda=\zeta=0$, i.e.\ $\vec{q}=\ep \vec{p}^{\,*}$
(a particular case of {\sl iv},   $N=0$ pure MQCD),  
$\Sigma$ is embedded in a four-dimensional manifold spanned by

$(x_4,x_5,x_6,x_{10})$. Finally if we set also $c=0$ besides
$\eta=\lambda=\zeta=0$ (an even more particular case of {\sl iv}), 
then $\Sigma$ is embedded in a three-dimensional space spanned by
$(x_4,x_5,x_{10})$ \refs\rwitqcd.

The family of $N=1$ curves parametrized by $\eta, \zeta$ 
(cases {\sl ii}\ and {\sl iii})
describes the soft breaking of the $N=2$ model to
$N=1$ by the mass of the adjoint superfield, i.e.\
\eqn\mad{ \Delta\CL_{N=1} = \int d^2 \theta \, m\,{\rm Tr} \Phi^2 \,+ 
\,{\rm h.c.}} 
Here, holomorphicity and global symmetries can be used to obtain a
precise match of the parameters appearing in the curve at weak string coupling
(recall that $R\sim (g_s)^{2/3}$) to the microscopic parameters of the 
effective low-energy field theory, i.e.\ the dynamical $\Lambda_{QCD}$ 
scales of the $N=2$ or $N=1$ models,  $\Lambda_2$, $\Lambda_1$,  and the
adjoint mass $m$ from \mad:  taking  
$v$ and $w$ with mass dimension, we have (see \refs\rmqcd) 
$\eta = (\Lambda_2)^2$, $ \zeta =
C_{\zeta}\,
\ell_{P}^2 (\Lambda_1)^3 /R \equiv \mu \,\eta$. With the one-loop matching
 $(\Lambda_1)^3 = 
m\,(\Lambda_2)^2$, we are led to $\mu =C_{\zeta}\, \ell_{P}^2 \,m/R$ and
$\eta=(\Lambda_1)^3/m$.
Notice that, with these matchings, the $N=2$ limit of the curves, eq.\ 
\cuno, does not exhibit explicit dependence on the compact radius $R$. This
agrees with the expectations from $N=2$ supersymmetry: all data encoded 
in the Seiberg--Witten
curve become protected against variations of
the string coupling when $\Lambda_2$ is kept fixed. 

In these $N=1$  models,  
both the microscopic theta angle
$\theta = n \,{\rm arg} (\eta)$, and the phase of the mass of the 
adjoint chiral multiplet $\alpha_m = {\rm arg} (m)$, are  
physically irrelevant, as one may absorb them into  phase redefinitions  
of $v,w,t$, which are allowed isometries of the background metric of
 eq.\ \bmm.  
In the field theory language, this is related to the existence of anomalous
$U(1)$ symmetries.  In addition, there are $n$ curves solving  
the ``field equations" eq.\ \emt\  for each
set of asymptotic data fixed at infinity. These solutions are related
by $2\pi$ shifts of the microscopic theta angle: a redefinition 
$(\eta, \zeta) \rightarrow (e^{i\delta} \eta, e^{i\delta} \zeta)$ can
be absorbed at infinity (i.e. in the region $z\sim 0$), by a
 reparametrization  $z\rightarrow e^{i\delta} z$, leaving the 
embedding of the curve 
in target space completely invariant, precisely if $\delta = 2\pi k/n$,
for $k=0,...,n-1$. This degeneracy is related to the existence of 
a symmetry at infinity: $t\rightarrow t$, $v\rightarrow v$, $w \rightarrow
e^{2\pi ik/n} w$, which is broken at finite distances by the brane
configuration. So, we have the usual picture of spontaneous symmetry
breaking by the gaugino condensate. For each of the $n$ vacua we
have ${\rm arg} (\zeta_k) = {\rm arg}( \bra {\rm Tr} \lambda\lambda \ket_k ) 
= (\theta + 2\pi k)/n $. The fact that vacua related by a spontaneously
broken symmetry are physically equivalent is realized on the curve 
by the fact that the redefinition $(\eta_k, \zeta_k )\rightarrow 
(\eta_{k+1}, \zeta_{k+1})$ can be absorbed into a reparametrization
$z\rightarrow e^{i\pi/n} z$, plus an isometry of the target 
$(v,w,t) \rightarrow (e^{i\pi/n} v, e^{i\pi/n} w, -t)$.     
This ensures the physical equivalence of   
 the tree-level effective Lagrangians obtained by reducing the M-theory
on each of the $n$ five-brane geometries labelled by $k$. 

In summary, the asymptotic behaviour of the  $N=1$ curves \cdos\  
depends only on the values of   $\zeta^n$ 
and $\zeta/\eta$, leading to $n$ equivalent curves,  which
are each mapped  into the next one by the phase transformation
$\theta\rightarrow \theta+ 2\pi$. Each individual vacuum (curve) 
is mapped into itself by $\theta\rightarrow \theta +2\pi n$. Another 
interesting property of the family of curves \cdos\ is the 
emergence of accidental additional symmetries in the MQCD limit
$\mu\rightarrow \infty$, $\zeta $ fixed, i.e.\ curves \ctres. A
 new $U(1)$ symmetry $(v,w,t)\rightarrow
(e^{i\delta} v, e^{-i\delta} w, e^{in\delta} t)$
appears only when $\eta \rightarrow 0$. 
Notice that this symmetry is not even present 
at infinity for $\eta \neq 0$.  

After supersymmetry breaking, i.e.\ when $\ep \neq 0$, we lose the
holomorphy constraints both on the geometry of the curve and on the
precise mapping between the microscopic parameters in the
effective low-energy, weak-coupling ($R \ll 1$) Lagrangian, 
and the parameters of the curve.
The only remaining
constraints would follow from selection rules imposed by global 
symmetries. The family of $N=1$ configurations has two natural $U(1)$
symmetries: $U(1)_v = U(1)_R $, associated to rotations in the $v$-plane,
an anomalous R-symmetry in the field theory description, and 
$U(1)_w = U(1)_J $, the non-anomalous R-symmetry surviving the full
$SU(2)_R$ of the $N=2$ models. The charges of the relevant quantities
under $U(1)_v \times U(1)_w$ are $Q(v) = (2,0)$, $Q(w) = (0,2)$, 
$Q(\mu) = Q(m) = (-2,2)$, $Q(\eta) = (4,0)$, $Q(\zeta) = (2,2)$, 
with $R$ inert under the phase redefinitions. 

For a non-zero supersymmetry breaking parameter, the symmetry
mentioned above, $\lambda\rightarrow
\zeta$, $w\rightarrow {\overline w}$,  
implies the effective bound $|\zeta| \ge \sqrt{|\eta \ep|}$, which
translates in the bound on the adjoint mass parameter: $|\mu| \ge 
\sqrt{|\ep / \eta|}$. Thus the family of configurations we consider
has a natural built-in hierarchy of soft
breakings, since most of parameter space satisfies   
$|\mu|^2 \gg |\ep/\eta|$. The analysis of the curves in eq.\ \curve\
becomes more tractable when $|\mu|^2 \gg |\ep/\eta|$. Indeed in this
case the adjoint mass parameter $\mu$ can be unambiguously
associated to a term of the form \mad\ since, in units of the natural 
$N=2$ scale $\eta$, the $N=1$ SUSY breaking scale is much larger
than the $N=0$ SUSY breaking scale. Moreover, in this case 
the supersymmetry breaking effects 
at low energies in the effective field theory
must be dominated by the gaugino mass since the
adjoint chiral multiplet masses are much larger. 
As a result, the associated field theoretical
models are generically the ones in refs.\ \refs\ryale, 
where spurion superfields lie in $N=1$ multiplets.
So we choose to parametrize 
supersymmetry breaking in the microscopic effective Lagrangian
in terms of the operator $m_{\lambda} {\rm Tr} \lambda\lambda$.
These considerations do not hold when $|\mu|^2 \sim |\ep/\eta|$,  
since the mass splitting between the ($N=1$) vector superfield and 
the adjoint chiral superfield, i.e.\ the scale of the 
$N=2 \rightarrow N=1$ breaking, is of the same order 
as the supersymmetry breaking parameter and it is natural to 
expect an  ${\cal O}(1)$ mixing of eq.\ \mad\ with 
the supersymmetry breaking operators. In this
case we could have a supersymmetry breaking pattern of the type 
studied in refs.\ \refs\rlag, with a full $N=2$ multiplet of spurions;  
however,  the analysis, based on global $U(1)$ selection rules, that 
we will pursue, is expected to be even less powerful, 
owing to the significant operator
mixing expected at the microscopic level.  
Thus in the rest of the paper we will restrict
ourselves to the case $|\mu|^2 \gg |\ep/\eta|$.

Therefore, saturating the effects of supersymmetry breaking with a  
gaugino mass, selection rules from the   $U(1)_v \times U(1)_w$ symmetry 
are easily derived,   assigning the
charges $Q(m_{\lambda}) = (-2,-2)$. Since eq.\ \curve\ 
fixes the charge of $\ep$ to be $Q(\ep) = (0,0)$, the global continuous
symmetries fix the dependence on the various CP-violating phases to be
\eqn\efes{\eqalign{ \ep &= f_{\ep} (\xi) \cr
\eta &= |\eta_{\rm susy}| e^{i\theta /n}\, (1+f_{\eta} (\xi)) \cr
\zeta &= |\zeta_{\rm susy} |
 e^{i(\alpha_m + \theta /n)} \,(1+ f_{\zeta} (\xi)). }}  
In these equations, $\xi = e^{i\theta_{ph} /n}$, with $\theta_{ph} = 
\theta + n\,{\rm arg}(m) + n\,{\rm arg} (m_{\lambda})$ the physical
theta angle, is invariant under the anomalous $U(1)_R$ symmetry, i.e.\ it
is the CP-violation parameter that cannot be rotated away by means
of anomalous phase rotations.  The functions $f_{\ep,\eta,\zeta} (\xi)$
depend on any real combination of the couplings in the theory 
($|m|$, $|m_{\lambda}|$, and  $|\Lambda_2|$), as well as
the eleven-dimensional Planck length $\ell_P$, and the 
compact radius $R$, 
and admit  power expansions
in the breaking parameter $m_{\lambda}$, such that they vanish in the
supersymmetric limit $m_\lambda\rightarrow 0$.

In fact, the generalized functions $f(\xi)$ are real functions, 
in the sense that, under complex conjugation, ${\overline {f(\xi)}}
 = f({\overline \xi})$.
This is due to the fact that a CP transformation acts on the curve by 
complex conjugation,\foot{This can be seen by recalling the form of the
generic $N=2$ curve,
where the $v$ plane is in fact the $\bra {\rm Tr} \Phi^2 \ket$
plane of the effective field theory. Also, in the $N=1$ models, a distance in
the $w$ plane is related to the expectation value of the dyons (see
\refs\rozdb), so CP really
acts by complex conjugation on the curve.}
but in the microscopic Lagrangian at weak string
coupling, it acts simply by inverting all CP-violating phases. So, to 
the extent that eq.\ \efes\ is a smooth limit of infrared quantities defined
at weak coupling, we should be able to characterize completely the CP
transformation in terms of complex conjugation of microscopic phases.

This structure with respect to CP-violating phases can be used to get
some insights in the theta-dependence of the $N=0$ MQCD theory described 
by the curve of eq.\ \curve. We have seen that, in  the
supersymmetric limit, $\ep=0=c^2 -1$, shifts of the theta angle by
$2\pi$ lead to identical boundary conditions at infinity,  leading  
to the appearance of $n$ degenerate 
 vacua, whose curves $C_k$ 
 are obtained by the replacement $\theta \rightarrow
\theta_k = \theta + 2\pi k$, with $k=0,...,n-1$.  
Once we break supersymmetry, the $n$-fold degeneracy of the vacuum is
lost. In field theory, a small supersymmetry breaking makes  
 $n-1$ of the vacua metastable, and  
 only one stable vacuum remains for generic values of the parameters. 
In the M-theory picture, we see that, as 
long as some of the functions in eq.\ \efes\ are non-trivial, 
the value of $\theta$ has
physical effects, as it cannot be rotated away into an isometry. This
is because the charge of all functions $f$ 
is trivial: $Q(f) =(0,0)$.
A stronger statement is that, because of 
the structure of the third equation in  \curve\ for $c^2 \neq 1$, 
no rescaling of $\ep$ can be absorbed into an isometry, 
and therefore the first equation in  \curve\ implies that no 
rescaling of $\eta$ is allowed either.

Thus, as we turn on a non-zero gaugino mass, each of the curves
$C_k$ is deformed, and the associated parameters $(\eta_k, \zeta_k,
\ep_k)$ are given by eq.\ \efes, with the substitutions 
$\theta\rightarrow \theta_k = \theta + 2\pi k$. The resulting $n$ vacua are
no longer physically equivalent  since, for generic values of
the parameters, the transformation $\theta\rightarrow \theta + 2\pi$
is not equivalent to a phase redefinition of $(\eta, \zeta)$, and
therefore it cannot be absorbed into a background isometry. One cannot
see directly in the M-theory picture the metastability of most of
the vacua, because the supergravity equations for the five-brane
only probe perturbative stability. Still, the euclidean
bounce solution leading to  a false vacuum decay via tunneling, could
perhaps appear as a particular interpolating five-brane configuration,
in analogy with the domain wall construction of ref. \ 
\refs\rwitqcd.~\foot{We
thank C. Bachas and M. Douglas for a discussion on this point.}

Now we face an apparent problem. The $\theta$ angle  now has physical
effects, but the  parameters of each curve $C_k$ 
are  periodic in $\theta$ with period $2\pi n$ instead of $2\pi$.
Still we know that the physics should be invariant under a 
$2\pi$ periodicity of $\theta$, and not $2\pi n$.
One might try to resolve the puzzle by invoking a special form of
the functions $f_{\eta,\zeta,\ep} (\xi)$ in \efes. For example, if
all supersymmetry breaking deformations are really functions of
$\xi^n$, then the parameters $(\eta_k, \zeta_k, \ep_k)$ become 
$2\pi$-periodic in $\theta$ up to a phase, which could be absorbed
into a background isometry. Such a dependence could be justified
by requiring that instantons completely saturate the supersymmetry
breaking deformations. However, this does not seem very likely,
because we have non-trivial branching in $\theta$ in the 
gaugino condensate already in the supersymmetric case. 
In addition, the analysis of field theory models of soft breaking
\refs\rtheta\ indicates that the puzzle of the ``wrong" theta
periodicity is resolved through a non-trivial spectral flow. That
is,  when $\theta\rightarrow\theta +2\pi$, a metastable 
vacuum will become stable and take the place of the original one. 
Thus at a particular value of $\theta$ there must be a level crossing of
two contiguous vacua, which are related by the redefinition 
$\theta\rightarrow \theta + 2\pi$. At this particular value of $\theta$
we should then find a spontaneously broken ${\bf Z}_2$ symmetry. 
According to standard lore, such a ${\bf Z}_2$ group is the action
of CP at that particular value of $\theta$, i.e. an example of the
Dashen phenomenon \refs\rdashen.

We can then, even with our limited knowledge of the curve of 
 eq.\ \curve\ 
given by eq.\ \efes, try to see if a CP transformation, i.e.\ a complex
conjugation of the curve, is a symmetry of the curve for a particular value 
of $\theta=\theta_c$. 

Indeed let us consider two curves related by the redefinition 
$\theta \rightarrow \theta+2\pi$, for example the curves $C_0$ and $C_{n-1}$.  
 Now we will show that at 
$(\theta_{ph})_0 =\pi$, i.e.\ at 
$\theta_c = \pi - n({\rm arg} (m) + {\rm arg} (m_{\lambda}))$, 
there is a symmetry between the $C_0$ and $C_{n-1}$ curves under a complex
conjugation (CP) and a background isometry. Indeed since
$\theta_k=\theta+2\pi k$,  $(\theta_{ph})_{n-1}= 2\pi n -\pi \simeq -\pi$
at the point where  $(\theta_{ph})_0 =\pi$. This implies that at 
$\theta=\theta_c$,  ${\overline \xi}_{n-1} = \xi_0$,  
and   $\epb_{n-1} = \ep_0$,
 ${\overline \eta}_{n-1} = e^{i\alpha_{\eta}} \,\eta_0$, 
$ {\overline \zeta}_{n-1} = 
e^{i\alpha_{\zeta}} \,\zeta_0$, with $\alpha_{\eta} = 2\,{\rm arg}(m) + 
2\,{\rm arg} (m_{\lambda})$, and 
$\alpha_{\zeta} = 2\, {\rm arg}(m_{\lambda})$.   
It is now trivial to check that such transformations can be absorbed into
an isometry of the $(v,w,t)$ space consisting of complex conjugation and
a phase redefinition, 
$(v,w,t)\rightarrow  (e^{-i\alpha_v} {\overline v}, e^{-i\alpha_w}
{\overline w},e^{-i\alpha_t} {\overline t})$, 
 with the values:\foot{Here we consider the physical situation with
${\rm Im}(c) =0$. It is only for this case that the $t$-rescaling corresponds
to a $U(1)_R$ transformation.}  
\eqn\values{\eqalign{
\alpha_v =& {\alpha_{\eta} \over 2} = {\rm arg}(m) + {\rm arg}(m_{\lambda}) \cr
\alpha_w =& \alpha_{\zeta}- {\alpha_{\eta} \over 2} = {\rm arg}(m_{\lambda}) 
- {\rm arg} (m) \cr 
\alpha_t =& \alpha_v \,n = n\,{\rm arg}(m) + n\,{\rm arg}(m_{\lambda}) }}    
This implies that, according to
 our interpretation \efes\ of the data of the curve 
in eq.\ \curve, at $(\theta_{ph})_0 =\pi$, i.e.\ at 
$\theta_c = \pi - n({\rm arg} (m) + {\rm arg} (m_{\lambda}))$,
there is a level crossing between the two vacua described by the curves
$C_0$ and $C_{n-1}$.  Clearly, such level crossings occur for any pair of
adjacent curves, so we can label as $C_0$ the absolutely stable curve at 
$(\theta_{ph})_0 =0$, without any loss of generality. 
{}From the known behaviour of the softly broken $N=1$ 
SQCD theories in field theory \refs\rtheta, we would then infer that for 
$-\pi < (\theta_{ph})_0 < \pi$ the $C_0$ curve describes the stable vacuum,
but that for $\pi < (\theta_{ph})_0 < 3\pi$ the stable vacuum is 
described by the $C_{n-1}$ curve,  while the $C_0$ curve now describes
a metastable vacuum. This picture thus reconciles the $2\pi n$ periodicity
in $\theta$ of each single curve, with the $2\pi$ periodicity in $\theta$
of the physics described by the $N=0$ MQCD curves  of eq.\ \curve.

To conclude, we would like to comment on the physical interpretation
of the discontinuous embeddings, i.e.\ the case of general complex $c$.
The relation between the curves parameters and the microscopic couplings
is encoded in the specific form of the functions in \efes. To the extent
that $\ep$ is real, we can interpret supersymmetry breaking in terms
of a continuous rotation of the brane configuration. However, one cannot
exclude that, for sufficiently strong supersymmetry breaking, the function
$f_{\ep} (\xi)$ develops an imaginary part at some point.  
This could be the geometrical interpretation of a phenomenon characteristic
of softly broken models with $N=2$ spurions \refs\rlag. In these models
one finds a number of metastable vacua of order $n$ in the large-$n$ limit,
but a particular metastable vacuum could disappear when transported 
around by a shift of $\theta$. Typically, the absolute vacuum at, say
$\theta =0$, becomes metastable at about $\theta =\pi$, and disappears
completely at $\theta \sim n\pi$. We can understand this phenomenon in
simple terms in 
a model with hierarchical supersymmetry breaking of the type discussed
in \refs\ryale\ and \refs\rtheta. The potential of the $N=1$ model
around a particular vacuum with condensation of $n-1$ dyons has the
structure $V(U) = m(\Lambda_2)^3 f(\sqrt{U} /\Lambda_2)$, with $U =
\bra {\rm Tr} \phi^2 \ket$, and $f(x)$ a function with  ${\cal O}(1)$
coefficients in the series expansion. Defining the 
dimensionless axion field as $a={\rm arg} (U)/2$, 
we have an axion potential with global scale
of order $m(\Lambda_2)^3$, and period $2\pi$. The leading-order correction in
the gaugino mass is of the form $\delta V \sim m_{\lambda} \bra 
{\rm Tr} \lambda \lambda \ket \sim m_{\lambda} m (\Lambda_2)^2 \,
e^{ia/n} $. Therefore, we see
that the local minimum of the axion at $\bra a \ket \sim  n\pi$
could be upset by the correction if $m_{\lambda} \sim \Lambda_2$.
If this happens,  the  axion's vacuum expectation value
at the local minimum becomes complex, and we interpret this as the 
disappearance of this vacuum at $\theta \sim n \pi$.

The previous discussion suggests that a complex value of $\ep$ as
a function of the microscopic parameters could be interpreted as
the disappearance of a metastable vacuum.  However, this phenomenon
occurs at $\theta \sim n\pi$, and thus      the picture
of level crossing at $\theta =\pi$ presented here remains,
at least for large enough $n$.   The general conclusion that
we can draw from this discussion is the robustness of the
level-crossing solution to the ``theta puzzle". Here we have
derived it from very general geometric considerations involving
no detailed analysis of dynamics.

\newsec{Acknowledgements}
We are indebted to  L. Alvarez-Gaum\'e, C. Bachas and M. Douglas
 for discussions and comments. 
This work is partially supported by the European Commission TMR programme
ERBFMRX-CT96-0045 in which A.P.\ is associated to the Institute for 
Theoretical Physics, K.U.\ Leuven. A.P. would like to thank CERN for its
hospitality while part of this work was carried out. 

\listrefs
\vfill\eject
\bye